# SALIENCY PREDICTION FOR OMNIDIRECTIONAL IMAGES CONSIDERING OPTIMIZATION ON SPHERE DOMAIN

*Bhishma Dedhia[1], Jui-Chiu Chiang[2] and Yi-Fan Char[2]*

[1]Department of Electrical Engineering, Indian Institute of Technology, Bombay, India
[2]Department of Electrical Engineering, National Chung Cheng University, Taiwan

**ABSTRACT**

There are several formats to describe the omnidirectional images. Among them, equirectangular projection (ERP), represented as 2D image, is the most widely used format. There exist many outstanding methods capable of well predicting the saliency maps for the conventional 2D images. But these works cannot be directly extended to predict the saliency map of the ERP image, since the content on ERP is not for direct display. Instead, the viewport image on demand is generated after converting the ERP image to the sphere domain, followed by rectilinear projection. In this paper, we propose a model to predict the saliency maps of the ERP images using existing saliency predictors for the 2D image. Some pre-processing and post-processing are used to manage the problem mentioned above. In particular, a smoothing based optimization is realized on the sphere domain. A public dataset of omnidirectional images is used to perform all the experiments and competitive results are achieved.

*Index Terms*—Omnidirectional image, ERP, saliency map, sphere

## 1. INTRODUCTION

In recent years, omnidirectional images (OMIs) have gained popularity owing to the increased demand of free-view navigation, particularly for the virtual reality (VR) applications. User can produce OMIs with several kinds of electronic devices and view OMIs on a head-mounted display (HMD). Visual exploration of OMIs is drastically different from that of the conventional images. Particularly, a much higher degrees of freedom of viewpoint is offered by OMIS. Consequently, it leads to enhanced interaction between the viewer and the scene, and an immersive entertainment is achieved accordingly.

Usually, the viewer will be interested in specified parts of the omnidirectional scene and it will bring benefit if the visual attention can be predicted. Being able to generate an accurate attention model is crucial in several computer vision related applications, such as action recognition [1], object tracking, and efficient compression specified for virtual and augmented reality [2, 3]. For example, in video streaming, it is essential to guarantee a real-time transmission under limited bandwidth. Since the data amount of OMI is quite huge, it will be wasteful to allocate the same resource for each part of OMIs. If the viewport image demanded by the viewer can be predicted, it is possible to assign more bits to the predicted region and less bits for the remaining parts during the encoding of the OMIs. With the guidance of the saliency map, it is feasible to make the delivery of viewport- on-demand more efficient.

Many remarkable works about saliency prediction of the conventional 2D image can be found in the literature [4-7]. Although the ERP image is represented as 2D image, these saliency predictors cannot perform well on it if no modification is realized. The ERP image suffers from the geometric distortion, which is propositional to the latitude. This issue should be addressed during the development of saliency prediction for the ERP image.

Since the ERP is one kind of projections of the OMIs, one solution to resolve the problem of geometric distortion is to represent the polar region in another format, such as cube map face, which is generated by cube map projection (CMP) [8-13]. Then, the saliency predictor for the conventional 2D image can be used to derive the visual attention of these cube faces, followed by re-projection to the ERP image. In [8], GVBS360 and BMS360 are proposed, which extend saliency models, namely Graph-based Visual Saliency (GVBS) [4] and Boolean Map Saliency (BMS) [5], designed for 2D image to the ERP image. Multi-plane projection is realized in [9] to simulate the viewing behavior of human eyes in HMDs. Then both the bottom-up and the top-down feature extractions are performed on each plane. The work in [10] uses fine-tuned SalGAN [6] for two image sources, including the original ERP and the cube faces images under several orientations. Then a fusion process is realized to generate the final saliency map in the ERP format. ERP images centered on two different longitude along the equator and cube map faces generated by rotating the cube center to several angles are used in [11] to generate the saliency map. In [13], the ERP image is split into patches and the sparse feature is extracted and an integrated saliency map is produced after taking the visual acuity and latitude bias into consideration.

Inspired by the work in [11], we use the ERP and the CMP image as the input and realize the saliency prediction individually. Since the spatial change of the visual attention is usually progressive, we realize an optimization on the saliency map to ensure a spatially edge-preserving smoothing.

In addition, eight ERP images, instead of 2, by rotating in the Y axis are generated to better manage the border artifact. The proposed approach demonstrates impressive results on the dataset of omnidirectional images.

## 2. METHODOLOGY

Omnidirectional images present the scene in a wider range, compared to the conventional image. However, not all of the areas of the omnidirectional image received intensive attention. Certain features of the image or the position on the sphere domain can be exploited to generate an accurate saliency map without the need for any training.

**2.1 Architecture of the proposed model**

In addition to the ERP, the cube map projection (CMP) which uses six square faces to present the surface of the sphere is also a popular format of OMIs. In the CMP, each face presents a 2D image for a particular viewport with a field of view (FOV) 90º and the conventional 2D saliency predictors can be applied to it with ease. Similar to the work [11], the proposed saliency map prediction computes both the saliency map for the ERP image and the CMP images and some modifications are proposed to improve the performance. Fig. 1 shows the overall architecture of the proposed saliency prediction model, which can be described by four main steps as follows:

(a) ERP-based saliency prediction
(b) CMP-based saliency prediction
(c) Introducing the equator bias
(d) Optimization on sphere domain

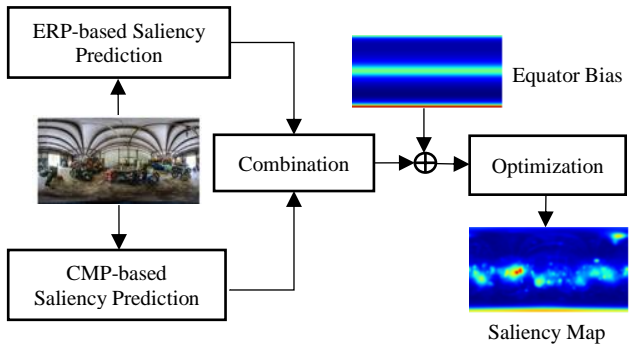

Fig. 1. Architecture of the proposed model

The saliency map for the OMI is parallelly predicted using steps (a) and (b). Saliency prediction for each cube face is yielded by BMS [5] while the saliency prediction for the ERP image employs SAM-Resnet [7]. Then two saliency models are integrated into one saliency map. Since many subjective experiments report that the content near the equator receives much higher attention, the equator bias map is introduced to mimic the viewers' behavior. Hence, an equator bias map is weighed into the combined saliency map. Then the map is smoothened on the sphere domain using an optimization based formulation. In the following, each step is introduced with details.

**2.2 ERP-based Saliency Prediction**

To preserve the saliency in the global context, more than one ERP images are used by changing the longitude of the image center along the equator. Different from [11], which considers 2 orientations, we have 8 orientations along the equator with a sampling rate of 45 degrees in the longitude. As mentioned before, the region near the equator in the ERP image preserves good geometry, while other regions suffer from the geometrical distortion, which is proportional to the latitude. It implies the conventional saliency predictors probably yield accurate attention models only in the middle portion. Thus, the idea in this approach is to predict the saliency of the middle portion and the edges for each ERP image separately. The edge portion is the region corresponds to the top and bottom faces of CMP, while the middle portion denotes the remaining region of the ERP, as illustrated in Fig. 2. Then, the edge portion will be represented by two face images.

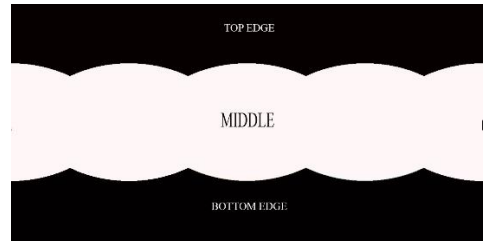

Fig. 2. ERP image is split into two edges and one middle portion

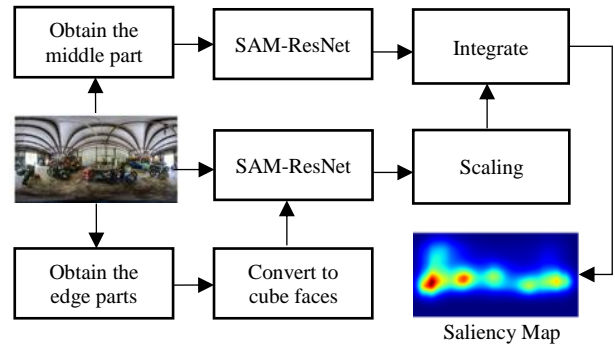

Fig. 3. Procedures of ERP-based saliency prediction

Saliency map for the middle portion and the two edge portions are generated directly by SAM-ResNet [7], which achieves good performance for the conventional image, as reported by the MIT300 benchmark [14,15]. Since the saliency maps of the cube faces under different orientation with fixed latitude can be seen as one map under varying rotations, there is no need to perform the saliency prediction for each orientation. Instead, only the saliency maps of the top and bottom faces of for the original orientation are predicted. For the middle portion, saliency map is generated for each orientation and these maps are fused into one map by taking the maximum value among them. Then the saliency of the middle portion and the edge portions will be assembled by projecting the top and bottom faces of the CMP to the ERP format. However, before the integration, the saliency map of the edge portion is scaled appropriately so that the maximum values in the middle portion and the edges are same. These procedures are illustrated in Fig. 3.

**2.3 CMP-based Saliency Prediction**

ERP images suffer from projection distortions, especially in the polar area. The six faces of CMP does not have this problem and conventional saliency predictors can be used. This has also been explored in [10,11].

BMS [5] is used to predict the saliency for each face. It is among the top scoring saliency prediction models in the MIT300 Benchmark [14,15]. The saliency map for each face will be stitched and re-projected back into the ERP format. However, this straightforward method leads to artifacts in the assembled saliency image. First, the CMP image is produced by rectilinear projection and the surface of the sphere is not uniformly projected to the cube face. It means that the boundary of each face will have larger distortion, compared to the center of the face. Besides, the discontinuity across the boundary of connected faces results in border effect. To address these issues and maintain the global context of the scene at the same time, we rotate the sphere with different orientation by changing the rotation angle of three axes. Five directions have been considered in total, by setting the rotating angles for axes (X, Y, Z) as (0°, 0°, 0°), (45°, 0°, 0°), (0°, 45°, 0°), (0°, 0°, 45°) and (45°, 45°, 0°). This procedure has also been also used in [11]. For each orientation, the saliency maps of the cube faces are re-projected to the ERP at the original orientation. Then a saliency map in the ERP format is obtained by taking the mean value of the saliency maps produced for all the orientations.

**2.3 Combining the Saliency Maps**

After obtaining the saliency for the ERP and the CMP, we will assemble them into a refined saliency map. Before combining them by averaging, the maps are scaled so that their maximum values are the same.

**2.4 Introducing the Equator Bias**

Since the regions near the equator are statistically attractive regions during the VR navigation, it is imperative to introduce the equator bias during the saliency prediction of the OMIs. The dataset [16] allows us to extract a global latitude-wise subjective attention map and the result is independent of the image characteristics. This latitude driven characterization will be considered to refine the saliency map generated in the previous processes. The equator-bias guided saliency at latitude $i$ is computed by (1):

$$E(i) = \frac{1}{m \times n} \sum_{p=1}^{n} \sum_{j=1}^{m} S_p(i,j), \quad (1)$$

where $S_p(i,j)$ denotes the subjective saliency value of the image $p$ at location $(i,j)$, and $n$ and $m$ denote the image numbers and the width of the image, respectively. Then a weighted average of the equator bias map, denoted as $S_{EB}$, and the saliency map, denoted as $S_{ini}$, generated from the previous steps, is fused. This procedure is illustrated in (2).

$$S_E = w \times S_{ini} + (1-w) \times S_{EB}, \quad (2)$$

where $w$ is empirically selected as 0.7 considering the contribution of the scene dependent characteristics and the equator bias.

**2.5 Optimization on Sphere Domain**

The last step involves smoothening the saliency map to remove the noise while maintaining the edge. An optimization-based approach is used to perform the task. The objective cost function is expressed in (3).

$$J(S_F) = \sum_p \left( \left(S_F^p - S_T^p\right)^2 + \lambda \sum_{q \in N(p)} \left(S_F^p - S_F^q\right)^2 \right), \quad (3)$$

where $S_F$ is the smoothened saliency map. $p$ and $q$ denote some specified pixels on $S_F$, respectively. $N(p)$ is the set of four nearest neighbors of a pixel $p$. $S_T$ is a manipulated version of $S_E$ through a masking operation. It means that the value of some pixels of $S_E$ is retained on $S_T$, while the remaining is set to 0. The mask is generated by considering the uniform sampling on the sphere surface and a spiral-based method [17] is adopted. In $S_T$, only the pixel, which corresponds to a uniformly sampled point on the sphere is preserved. The reason behind is that neighboring pixels in ERP format do not have fixed distance in the sphere domain and not all the pixels in the ERP domain have equal importance. Similar to the metric of S-PSNR [18], which computes the PSNR on selected pixels, which are uniformly distributed on a sphere surface, we select the uniformly sampled pixels on the sphere and project them back to the ERP image to form the mask. These pixels are served as seeds and the smoothing is realized. The number of points sampled on the sphere surface was directly proportional to the size of the ERP image, and the sampling number per steradian, denotes as $N_s$, is defined in (4),

$$N_s = H \times W \times K \quad (4)$$

where $H$ and $W$ denote the height and width of the input ERP image, respectively. $K$ is a parameter and it determines the density of the samples. We vary the parameter $K$ as 10, 100 and 1000. Table 1 summarizes the scores of the test image P33 [19] for various values of $K$. It indicates that a remarkable improvement is achieved after performing the optimization on the sphere domain, in particular for the Normalized Saliency Scanpath (NSS) metric [20].

Table 1. Performance for various parameters $K$ for test image P33

|  | without optimization | K=10 | K=100 | K=1000 |
|---|---|---|---|---|
| KLD↓ | 0.46 | 0.40 | 0.39 | 0.39 |
| CC↑ | 0.67 | 0.71 | 0.70 | 0.70 |
| NSS↑ | 0.53 | 0.96 | 0.96 | 0.96 |
| AUC↑ | 0.58 | 0.69 | 0.69 | 0.69 |

## 3. EXPERIMENTAL RESULTS

The proposed scheme is first evaluated over the database for the omnidirectional images [19], which includes the original images and both the head movement and head-eye movement after conducting the subjective experiments. Then the head saliency and head-eye saliency are provided and served as ground truth. The head-eye movement is considered in this work. In the beginning, we present the results of the CMP-based saliency model. Four popular objective metrics in the saliency community are used, including Kullback-Leibler Divergence (KLD), Pearson's Correlation Coefficient (CC), Normalized Saliency Scanpath (NSS) and AUC-Judd [20]. The toolbox [21] is used to compute these scores. Table 2 reports the performance of the CMP-based saliency prediction for the test image P33. A significant improvement in terms of KLD is achieved when different orientations are employed.

Table 2. Score for CMP-based saliency prediction

|  | CMP (one orientation) | CMP (five orientations) |
|---|---|---|
| KLD | 3.50 | 0.56 |
| CC | 0.42 | 0.51 |
| NSS | 0.58 | 0.58 |
| AUC | 0.59 | 0.62 |

Table 3 shows the results when the saliency maps of the ERP-based and the CMP-based are combined. Not surprising, a better performance is achieved in this way.

Table 3. Score for the combined saliency map

|  | ERP-based saliency map | Combined with CMP-based saliency map |
|---|---|---|
| KLD | 0.55 | 0.39 |
| CC | 0.56 | 0.69 |
| NSS | 0.51 | 0.65 |
| AUC | 0.66 | 0.62 |

Fig. 4 presents the saliency map of the test image P33 when the optimization is performed. It shows that the optimization on the sphere domain indeed improves the saliency map. In particular, some regions on the right side of the image have a larger saliency value after performing the optimization, which is consistent with the subjective behavior.

Table 4 shows the result of the proposed work for the dataset [19], while Table 5 summarizes the performance for the dataset [22], which is the verification dataset for ICME 2017 Grand Challenge Salient360! Several works are compared in this table. It shows that all the model has a similar AUC score. The proposed technique outperforms the other schemes in NSS score and it achieves a comparable performance in terms of KLD and CC metrics.

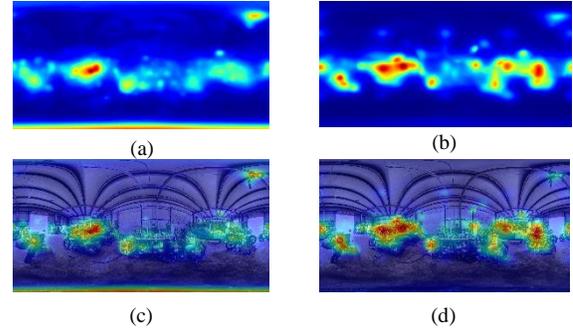

(a)  (b)

(c)  (d)

Fig. 4. Saliency map of the test image P33 [19]. (a) Without optimization (b) With optimization, $K$=10 (c) input image overlaid with (a), (d) input image overlaid with (b)

Table 4. Results for the head-eye movement prediction with dataset [19]

| Method | KLD | CC | NSS | AUC |
|---|---|---|---|---|
| Proposed | 0.515 | 0.545 | **1.002** | 0.719 |

Table 5. Results for the head-eye movement prediction with dataset [22]

| Method | KLD | CC | NSS | AUC |
|---|---|---|---|---|
| GVBS360[8] | 0.698 | 0.527 | 0.851 | 0.714 |
| [9] | 0.481 | 0.532 | 0.918 | 0.734 |
| [10] | 0.431 | **0.659** | 0.971 | **0.746** |
| [11] | **0.42** | 0.61 | 0.81 | 0.72 |
| [13] | 0.477 | 0.550 | 0.936 | 0.736 |
| Proposed | 0.469 | 0.570 | **1.027** | 0.731 |

## 4. CONCLUSION

This work proposes techniques to predict the saliency map of the omnidirectional images based on saliency predictor of the conventional 2D image. Both ERP-based and CMP-based saliency prediction is realized. The experimental results show that an average model is superior in performance over the individual CMP-based model and ERP-based model. The optimization on the sphere domain brings additional improvement. The proposed techniques have considerably good performance in predicting the NSS and AUC. Future work can consider several changes to improve this model. The ERP-based saliency prediction can be replaced by multi-view based saliency prediction to better extract the visual attention. Moreover, this model can be combined with a training-based saliency predictor to yield improved results.